\documentstyle[12pt,a4]{article}

\begin{document}

\newcommand{\nc}[2]{\newcommand{#1}{#2}}
\newcommand{\ncx}[3]{\newcommand{#1}[#2]{#3}}
\ncx{\pr}{1}{#1^{\prime}}
\nc{\nl}{\newline}
\nc{\np}{\newpage}
\nc{\nit}{\noindent}
\nc{\vs}{\vspace{2ex}}
\nc{\be}{\begin{equation}}
\nc{\ee}{\end{equation}}
\nc{\ba}{\begin{array}}
\nc{\ea}{\end{array}}
\nc{\bea}{\begin{eqnarray}}
\nc{\eea}{\end{eqnarray}}
\nc{\nb}{\nonumber}
\nc{\dsp}{\displaystyle}
\nc{\bit}{\bibitem}
\nc{\ct}{\cite}
\ncx{\dd}{2}{\frac{\partial #1}{\partial #2}}
\nc{\pl}{\partial}
\nc{\dg}{\dagger}
\nc{\wig}{\wedge}
\nc{\cL}{{\cal L}}
\nc{\cD}{{\cal D}}
\nc{\cF}{{\cal F}}
\nc{\cG}{{\cal G}}
\nc{\cJ}{{\cal J}}
\nc{\cQ}{{\cal Q}}
\nc{\tB}{\tilde{B}}
\nc{\tD}{\tilde{D}}
\nc{\tF}{\tilde{F}}
\nc{\tH}{\tilde{H}}
\nc{\tQ}{\tilde{Q}}
\nc{\tR}{\tilde{R}}
\nc{\tZ}{\tilde{Z}}
\nc{\tg}{\tilde{g}}
\nc{\tog}{\tilde{\og}}
\nc{\tGam}{\tilde{\Gam}}
\nc{\tPi}{\tilde{\Pi}}
\nc{\tcD}{\tilde{\cD}}
\nc{\tcQ}{\tilde{\cQ}}
\nc{\ag}{\alpha}
\nc{\bg}{\beta}
\nc{\gam}{\gamma}
\nc{\Gam}{\Gamma}
\nc{\bgm}{\bar{\gam}}
\nc{\del}{\delta}
\nc{\Del}{\Delta}
\nc{\eps}{\epsilon}
\nc{\ve}{\varepsilon}
\nc{\zg}{\zeta}
\nc{\th}{\theta}
\nc{\vt}{\vartheta}
\nc{\Th}{\Theta}
\nc{\kg}{\kappa}
\nc{\lb}{\lambda}
\nc{\Lb}{\Lambda}
\nc{\ps}{\psi}
\nc{\Ps}{\Psi}
\nc{\sg}{\sigma}
\nc{\spr}{\pr{\sg}}
\nc{\Sg}{\Sigma}
\nc{\rg}{\rho}
\nc{\fg}{\phi}
\nc{\Fg}{\Phi}
\nc{\vf}{\varphi}
\nc{\og}{\omega}
\nc{\Og}{\Omega}
\nc{\Kq}{\mbox{$K(\vec{q},t|\pr{\vec{q}\,},\pr{t})$ }}
\nc{\Kp}{\mbox{$K(\vec{q},t|\pr{\vec{p}\,},\pr{t})$ }}
\nc{\vq}{\mbox{$\vec{q}$}}
\nc{\qp}{\mbox{$\pr{\vec{q}\,}$}}
\nc{\vp}{\mbox{$\vec{p}$}}
\nc{\va}{\mbox{$\vec{a}$}}
\nc{\vb}{\mbox{$\vec{b}$}}
\nc{\Ztwo}{\mbox{\sf Z}_{2}}
\nc{\Tr}{\mbox{Tr}}
\nc{\lh}{\left(}
\nc{\rh}{\right)}
\nc{\ld}{\left.}
\nc{\rd}{\right.}
\nc{\nil}{\emptyset}
\nc{\bor}{\overline}
\nc{\ha}{\hat{a}}
\nc{\da}{\hat{a}^{\dg}}
\nc{\hb}{\hat{b}}
\nc{\db}{\hat{b}^{\dg}}
\nc{\hN}{\hat{N}}
\ncx{\abs}{1}{\left| #1 \right|}

\pagestyle{empty}

\begin{flushright}
NIKHEF/95-061 \\
hep-th/9511166
\end{flushright}

\vspace{3ex}
\begin{center}
{\Large{\bf KILLING TENSORS AND A NEW }}\\
\vspace{2ex}

{\Large{\bf GEOMETRIC DUALITY }} \\
\vspace{5ex}

{\large R.H.\ Rietdijk$^*$ \\
Un.\ of Durham, U.K. }\\
\vspace{2ex}

{\large J.W.\ van Holten \\
NIKHEF, Amsterdam NL} \\
\vs

\date{\today}
\vspace{7ex}

{\small{\bf Abstract}}\\
\end{center}
\vspace{1ex}

\noindent
{\small
We present a theorem describing a dual relation between the local
geometry of a space admitting a symmetric second-rank Killing tensor, and
the local geometry of a space with a metric specified by this Killing tensor.
The relation can be generalized to spinning spaces, but only at the expense
of introducing torsion. This introduces new supersymmetries in their geometry.
Interesting examples in four dimensions include the Kerr-Newman metric of
spinning black-holes and self-dual Taub-NUT. }
\vfill

\nit
$\overline{\mbox{\footnotesize $^*$ Address since sept.\ 1, 1995:}}$ \nl
{\footnotesize Kon.\ Shell Research Lab., P.O.\ Box 60, 2280 AB Rijswijk NL}

\newpage

\pagestyle{plain}
\pagenumbering{arabic}

\section{Introduction}{\label{S.1}}

The importance of symmetries in the description of physical systems can
hardly be over-estimated. In the case of dynamical systems in particular,
continuous symmetries determine the structure of the algebra of observables
by Noether's theorem, giving rise to constants of motion in classical
mechanics and quantum numbers labeling stationary states in quantum theory.

In a geometrical setting symmetries are connected with isometries associated
with Killing vectors and, more generally, Killing tensors on the
configuration space of the system. An example is the motion of a point
particle in a space with isometries \ct{RvH1}, which is a physicist's way of
studying the geodesic structure of a manifold. Contact with the algebraic
approach is made through Lie-derivatives and their commutators.
In \ct{RvH1,RvH2} such studies were extended to spinning space-times
described by supersymmetric extensions of the geodesic motion, and in
\ct{RvH3} it was shown that this can give rise to interesting new
types of supersymmetry as well.

This paper concerns spaces on which there exists a symmetric second-rank
Killing tensor, including such interesting cases as the four-dimensional
Kerr-Newman space-time describing spinning and/or charged black holes, and
four-dimensional Taub-NUT which describes a gravitational instanton \ct{Haw}
and appears as a low-energy effective action for monopole scattering
\ct{Man,GMan}. These second-rank Killing tensors correspond to constants of
motion which are quadratic in the momenta and play an important role in the
complete solution of the problem of geodesic motion in these spaces \ct{BC1}.
Similar results concerning the monopole with spin and 3-D fermion systems
have been presented in \ct{M4,Plyush}.

The main aim of the paper is to present and illustrate a theorem concerning
the reciprocal relation between two local geometries described by metrics
which are Killing tensors with respect to one another. In sect.\ \ref{S.2}
the basic theorem is presented. It is illustrated in sect.\ \ref{S.3} with
the four-dimensional examples mentioned above. In sect.\ \ref{S.4} we
extend the discussion to the motion of particles with spin and charge
in curved space, including torsion and electro-magnetic background fields.
This generalizes results obtained by Tanimoto \ct{Tan}. In sect.\ \ref{S.5}
the relation between Killing tensors and new supersymmetries, representing a
certain square root of the associated constants of motion, is explained.
The basic ingredient is the existence of so-called Killing-Yano tensors
\ct{RvH3}. The duality theorem is then generalized to include geometries with
Killing-Yano tensors, and it is shown that in general this requires the
introduction of torsion. Finally the explicit expressions for the Killing-Yano
tensors and the associated torsion-tensors for the examples of Kerr-Newman and
Taub-NUT are given.

\section{Dual geometries}{\label{S.2}}

Suppose a space (of either Euclidean or Lorentzian signature) with metric
$g_{\mu\nu}(x)$ admits a second-rank Killing tensor field $K_{\mu\nu}(x)$:

\be
 K_{(\mu\nu;\lb)}\, =\, 0.
\label{1.1}
\ee

\nit
Here the semi-colon denotes a Riemannian covariant derivative, and the
parentheses denote complete symmetrization over the component indices.
The equation of motion of a particle on a geodesic is derived from the
action

\be
S\, =\,
    \int d\tau \lh \frac{1}{2}\, g_{\mu\nu} \dot{x}^{\mu} \dot{x}^{\nu} \rh.
\label{1.2}
\ee

\nit
The corresponding world-line hamiltonian is constructed in terms of the
inverse (contravariant) metric $g^{\mu\nu}$:

\be
H\, =\, \frac{1}{2}\, g^{\mu\nu} p_{\mu} p_{\nu},
\label{1.3}
\ee

\nit
and the elementary Poisson brackets are

\be
\left\{ x^{\mu}, p_{\nu} \right\}\, =\, \del^{\mu}_{\nu}.
\label{1.3.1}
\ee

\nit
The equation of motion for a phase space function $F(x,p)$ can be computed
from the Poisson brackets with the hamiltonian:

\be
\frac{dF}{d\tau}\, =\, \left\{ F, H \right\}.
\label{1.4}
\ee

\nit
{}From the contra-variant components $K^{\mu\nu}$ of the Killing tensor one can
construct a constant of motion $K$:

\be
K\, =\, \frac{1}{2}\, K^{\mu\nu} p_{\mu} p_{\nu}.
\label{1.5}
\ee

\nit
Its bracket with the hamiltonian vanishes precisely because of the condition
(\ref{1.1}). The constant of motion $K$ generates symmetry transformations
on the phase space linear in momentum:

\be
\left\{ x^{\mu}, K \right\}\, =\, K^{\mu\nu} p_{\nu}.
\label{1.5.1}
\ee

\nit
The infinitesimal transformation with parameter $\ag$ can also be written
in terms of the velocity as

\be
\del(\ag)\, x^{\mu}\, =\, \ag\, K^{\mu}_{\:\:\nu} \dot{x}^{\nu}.
\label{1.5.2}
\ee

\nit
The formal similarity between the constants of motion $H$ and $K$, and the
symmetrical nature of the condition implying the existence of the Killing
tensor:

\be
\left\{ H, K \right\}\, =\, 0,
\label{1.6}
\ee

\nit
amount to a reciprocal relation between two different models: the model
with hamiltonian $H$ and constant of motion $K$, and a model with constant
of motion $H$ and hamiltonian $K$. In the second model $H$ generates a
symmetry in the phase space of the system. Thus the relation between the
two models has a geometrical interpretation: it implies that if $K^{\mu\nu}$
are the contravariant components of a Killing tensor with respect to the
inverse metric $g^{\mu\nu}$, then $g^{\mu\nu}$ must represent a Killing tensor
with respect to the inverse metric defined by $K^{\mu\nu}$.

A more direct proof of this result can be given purely in terms of
geometrical quantities. Given the metric $g_{\mu\nu}$ and the Killing
tensor $K_{\mu\nu}$ satisfying eq.(\ref{1.1}), we identify the contravariant
components of $K$ with a new contravariant metric $\tilde{g}^{\mu\nu}$:

\be
\tilde{g}^{\mu\nu}\, \equiv\, K^{\mu\nu}\, =\, g^{\mu\lb} K_{\lb\kg}
g^{\kg \nu}.
\label{1.7}
\ee

\nit
If these components form a non-singular $(d \times d)$-dimensional matrix,
we denote its inverse by $\tilde{g}_{\mu\nu}$:

\be
\tilde{g}^{\mu\lb}\, \tilde{g}_{\lb\nu}\, =\, \del^{\mu}_{\nu}.
\label{1.8}
\ee

\nit
Now we can interpret $\tilde{g}_{\mu\nu}$ as the metric of another space.
Define the associated Riemann-Christoffel connection
$\tGam_{\mu\nu}^{\:\:\:\:\:\lb}$ as usual through the metric postulate

\be
\tD_{\lb} \tilde{g}_{\mu\nu}\, =\, 0.
\label{1.9}
\ee

\nit
Then it follows from eq.(\ref{1.1}) that

\be
\tD_{\lh\lb\right.} \tilde{K}_{\left. \mu\nu\rh}\, =\, 0,
\label{1.10}
\ee

\nit
where $\tilde{K}_{\mu\nu}$ are the covariant components of $g$ with respect to
the metric $\tg$:

\be
\tilde{K}_{\mu\nu}\, =\, \tg_{\mu\lb} g^{\lb\kg} \tg_{\kg\nu}.
\label{1.11}
\ee

\nit
This reciprocal relation between the metric structure and certain symmetries
of pairs of spaces therefore constitutes a duality relation: performing the
operation of mapping a Killing tensor to a metric twice leads back to the
original geometry.

\section{Examples: Kerr-Newman and Taub-NUT}{\label{S.3}}

Examples of manifolds with Killing-tensor fields include physically
important ones, like the four-dimensional Kerr-Newman and Taub-NUT
solutions of the Einstein or Einstein-Maxwell equations. In this section
we present explicit expressions for their metric, Killing tensor and the
metric of their dual spaces in the sense defined above.
\vspace{2ex}

\nit
{\em Kerr-Newman.} The Kerr-Newman geometry describes a charged spinning
black hole; in a standard choice of co-ordinates the metric is given by
the following line-element:

\be
\ba{lll}
\dsp{ds^2} & = &
\dsp{ - \, \frac{\Del}{\rg^2} \left[ dt - a \sin^2 \th \, d \vf \right]^2
+ \frac{\sin^2 \th}{\rg^2} \left[ (r^2 + a^2) d \vf - a dt \right]^2 + \nb} \\
 & & \\
 & & \dsp{ + \, \frac{\rg^2}{\Del} dr^2 + \rg^2 d\th^2 ,}
\ea
\label{1.12}
\ee

\nit
Here

\be
\ba{lll}
\dsp{\Del} & = & \dsp{r^2 + a^2 - 2Mr + Q^2 , \nb }\\
 & & \\
\dsp{\rg^2} & = & \dsp{r^2 + a^2 \cos^2 \th ,}
\ea
\label{1.13}
\ee

\nit
with $Q$ the background electric charge, and $J = Ma$ the total angular
momentum. The expression for $ds^2$ only describes the fields
{\em outside} the horizon, which is located at

\be
r = M + \sqrt{M^2 - Q^2 - a^2} .
\label{1.14}
\ee

\nit
The background electric charge also creates an electro-magnetic field
described by the Maxwell 2-form

\be
\ba{lll}
\dsp{F} & = & \dsp{\frac{Q}{\rg^4} (r^2 - a^2 \cos^2 \th) dr \wedge
\left[ dt - a \sin^2 \th \, d \vf \right] + \nb }\\
 & & \\
 & + &  \dsp{\frac{2Qar \cos \th \sin \th}{\rg^4} d\th \wedge
\left[ - a dt + (r^2 + a^2) d\vf  \right] .}
\label{1.15}
\ea
\ee

\nit
The Kerr-Newman metric admits a second-rank Killing tensor field, which can
be described in this co-ordinate system by the quadratic form

\be
\ba{lll}
K & = & K_{\mu\nu}\, dx^{\mu} dx^{\nu} \\
  &  & \\
  & = & \dsp{ \frac{a^2 \cos^2 \theta \Del}{\rg^2}
              \left[ dt - a \sin^2 \th \, d \vf \right]^2 +
              \frac{r^2 \sin^2 \th}{\rg^2} \left[ (r^2 + a^2) d \vf
              - a dt \right]^2 \nb} \\
 & & \\
 & & \dsp{ - \, \frac{\rg^2 a^2 \cos^2 \theta}{\Del} dr^2 +
           r^2 \rg^2  d\th^2}
\ea
\label{1.16}
\ee

\nit
Its contravariant components define the inverse metric $\tg^{\mu\nu}$ of the
dual geometry. The dual line-element then becomes

\be
\ba{lll}
\dsp{d \tilde{s}^2} & = &  \tg_{\mu\nu} dx^{\mu} dx^{\nu} \\
  & & \\
  & = & \dsp{ \frac{\Del}{\rg^2 a^2 \cos^2 \theta}
              \left[ dt - a \sin^2 \th \, d \vf \right]^2 +
              \frac{\sin^2 \th}{\rg^2 r^2} \left[ (r^2 + a^2) d \vf
              - a dt \right]^2 \nb} \\
 & & \\
 & & \dsp{ - \, \frac{\rg^2}{\Del a^2 \cos^2 \theta} dr^2 +
           \frac{\rg^2}{r^2} d\th^2} ,
\ea
\label{1.17}
\ee

\nit
providing an explicit expression for the dual metric.
\vspace{2ex}

\nit
{\em Taub-NUT.} The four-dimensional Taub-NUT metric depends on a parameter
$m$ which can be positive or negative, depending on the application; for
$m > 0$ it represents a non-singular solution of the self-dual Euclidean
Einstein equation and as such is interpreted as a gravitational instanton.
A standard form of the line-element is

\be
\ba{lll}
ds^2 & = & \dsp{ \lh 1 + \frac{2m}{r} \rh \lh dr^2 + r^2 d\th^2
                 + r^2 \sin^2 \th d\vf^2 \rh }\\
 & & \\
 & & \dsp{
     +\, \frac{4m^2}{1 + \frac{2m}{r}}\, \lh d\psi + \cos \th d\vf \rh^2. }
\ea
\label{1.18}
\ee

\nit
A symmetric Killing tensor with respect to this metric is represented by
the quadratic form

\be
\ba{lll}
K & = & \dsp{ \lh 1 + \frac{2m}{r} \rh \lh dr^2 +
 \frac{r^2}{m^2} \lh r + m \rh^2 \lh d\th^2 + \sin^2 \th d\vf^2 \rh\rh\, }\\
  & & \\
  & & \dsp{
      +\, \frac{4m^2}{1 + \frac{2m}{r}}\, \lh d\psi + \cos \th d\vf \rh^2. }
\ea
\label{1.19}
\ee

\nit
The matrix inverse of the contravariant form of $K$ gives the dual line
element

\be
\ba{lll}
d\tilde{s}^2 & = & \dsp{ \lh 1 + \frac{2m}{r} \rh \lh dr^2 +
 \frac{m^2r^2}{(r+m)^2} \lh d\th^2 + \sin^2 \th d\vf^2 \rh\rh\, }\\
  & & \\
  & & \dsp{
      +\, \frac{4m^2}{1 + \frac{2m}{r}}\, \lh d\psi + \cos \th d\vf \rh^2. }
\ea
\label{1.20}
\ee

\nit
The Taub-NUT metric admits three more second rank Killing tensors. They form a
conserved vector of the Runge-Lenz type and are given by

\be
\ba{lll}
K_{(i)} & = & \dsp{ - \frac{2}{m} \left( 1 + \frac{2m}{r} \right)
\left( 1 + \frac{m}{r} \right) r_{i}
\left( dr^2 + r^2 d \th^2 + r^2 \sin^2 \th d\vf^2 \right) }\\
 & & \\
 & & \dsp{ + \; \frac{8 m^2}{r \left( 1 + \frac{2m}{r} \right)} r_{i}
\left( d \psi + \cos \th d \vf \right)^2 +
\frac{2r}{m} \left( 1 + \frac{2m}{r} \right)^2 drdr_{i} }\\
 & & \\
 & & \dsp{ + \; 4 \left( 1 + \frac{2m}{r} \right)
\left( \vec{r} \times d \vec{r} \right)_{i}
\left( d \psi + \cos \th \vf \right). }
\ea
\label{1.21}
\ee

\nit
These conserved quantities define the dual line-elements

\be
\ba{lll}
d\tilde{s}^{2}_{(i)} & = & \dsp{ \frac{- 1}{r_{i}^{2} -\left(r+2m\right)^2}
\left\{ - \frac{2m^2}{r} \left( 1 + \frac{2m}{r} \right) r_{i}
\left( dr^2 + r^2 d \th^2 + r^2 \sin^2 \th d\vf^2 \right) \right.}\\
 & & \\
 & & \dsp{ + \; \frac{8 m^3 \left(1+\frac{m}{r} \right)}
{\left( 1 + \frac{2m}{r} \right)} r_{i}
\left( d \psi + \cos \th d \vf \right)^2 +
2mr \left( 1 + \frac{2m}{r} \right)^2 dr dr_{i} }\\
 & & \\
 & & \dsp{ \left. + \; 4 m^2 \left( 1 + \frac{2m}{r} \right)
\left( \vec{r} \times d \vec{r} \right)_{i}
\left( d \psi + \cos \th \vf \right) \right\}. }
\ea
\label{1.22}
\ee

\nit
Note that in all examples the dual metrics are very similar, though not
identical, to the original ones. The physical interpretation of the dual
metrics remains to be clarified.

\section{Electro-magnetism and torsion}{\label{S.4}}

The results obtained so far can be generalized to include spin and charge
for the test particle. Electro-magnetic interactions are introduced
via minimal coupling, whilst spin can be described by the supersymmetric
extension of the dynamics, using a vector of Grassmann-odd co-ordinates
$\psi^{\mu}$. This also allows for torsion to be present, as is
actually required for some of our purposes. As found in \ct{RvH3}, the
existence of Killing-tensors now requires new supersymmetric structures.
The inclusion of charge was studied in \ct{Tan}.

Using these procedures, consider a charged, spinning particle which moves
under the influence of an electro-magnetic field $A_{\mu}$, a gravitational
field $g_{\mu \nu}$, and a completely anti-symmetric torsion field $A_{abc}$,
as described by the action

\be
\ba{lll}
S & = & \dsp{\int d \tau \left\{
\frac{1}{2}\, g_{\mu \nu} \dot{x}^{\mu} \dot{x}^{\nu}
- \frac{i}{2}\, \eta_{a b}  \dot{\ps}^a \ps^b
- \frac{i}{2} \dot{x}^{\mu} \ps^a \ps^b
\left( \og_{\mu a b} + A_{\mu a b} \right) +
\right. } \\
& & \\
& & \dsp{\hspace{15mm} \left.
+ \; \dot{x}^{\mu} A_{\mu}
- \frac{i}{2} \ps^a \ps^b F_{a b}
- \frac{1}{4!} \ps^a \ps^b \ps^c \ps^d F_{a b c d}
\right\}} .
\ea
\label{2.1}
\ee

\nit
Here $\mu, \nu, \ldots$ label space-time coordinates, while $a,b,\ldots$ label
local Lorentz coordinates. Both types of indices run from $1, \ldots , d$, the
dimension of space-time. They can be converted into each other by contracting
with an (inverse) vielbein $e_{a}^{\;\; \mu}(x)$ ($e_{\mu}^{\;\; a}(x)$),
defined by

\be
e_{a}^{\;\; \mu} e_{b}^{\;\; \nu} \eta^{ab} = g^{\mu \nu} .
\label{2.2}
\ee

\nit
The torsion will usually be written explicitly, except in a few cases where
concise notation requires otherwise; this is always manifest in
the text and the equations. In any case, $\og_{\mu a b}$ is the bare spin
connection, defined by

\be
\og_{\mu a b} = e_{a}^{\;\; \kg} e_{b[\mu,\kg]} -
e_{b}^{\;\; \kg} e_{a[\mu,\kg]} -
e_{a}^{\;\; \kg} e_{b}^{\;\; \lb} e_{c \mu} e^{c}_{\; [\kg,\lb]}.
\label{2.3}
\ee

\nit
We will use square brackets to denote anti-symmetrisation and parentheses to
denote symmetrisation over all indices enclosed, with total weight equal to
one. Finally $F_{a b}$ and $F_{a b c d}$ are contractions with vielbeins of
the field strengths associated with $A_{\mu}$ and $A_{\mu \nu \kg}$

\be
\ba{lll}
F_{\mu \nu} & = & 2 \pl_{[\mu} A_{\nu]} , \\
& & \\
F_{\mu \nu \kg \lb} & = & 4 \pl_{[\mu} A_{\nu \kg \lb]}
\ea
\label{2.4}
\ee

\nit
The classical equations of motion derived from the action (\ref{2.1}) are

\be
g_{\mu\nu} \frac{D^2 x^{\nu}}{D\tau^2}\, =\, F_{\mu\nu} \dot{x}^{\nu}\,
  -\, \frac{i}{2} \psi^a \psi^b R_{ab\mu\nu}^{T} \dot{x}^{\nu}\, -\,
  \frac{i}{2} \psi^a \psi^b D_{\mu}^{T} F_{ab}\, -\, \frac{1}{4}\,
  \psi^a \psi^b \psi^c \psi^d D_{\mu}^{T} F_{abcd} ,
\label{2.4.1}
\ee

\nit
for the orbit of the particle, and

\be
\frac{D^{T}}{D\tau} \psi^a\, =\, F^{a}_{\:\:b} \psi^b\, -\,
  \frac{i}{3!} F^a_{\:\:bcd} \psi^a \psi^b \psi^c,
\label{2.5.1}
\ee

\nit
for the precession of the spin. In these equations we have by exception
included torsion in the derivatives, as denoted by the superscript $T$,
because this simplifies the expressions considerably.

The action (\ref{2.1}) is invariant under the supersymmetry transformation

\be
\ba{lll}
\del_{\eps} x^{\mu} & = & -i \eps \ps^a e_{a}^{\;\; \mu} , \\
& & \\
\del_{\eps} \ps^a & = & \eps \dot{x}^{\mu} e_{\mu}^{\;\; a}
+ i \eps \ps^b \ps^c e_{b}^{\;\; \mu} \og_{\mu c}^{\;\;\;\; a} .
\ea
\label{2.5}
\ee

\nit
This implies that there is a conserved quantity $Q$, the supercharge, which
generates the transformation by its Dirac-bracket. The fundamental
Dirac-brackts are

\be
\left\{ x^{\mu} , P_{\nu} \right\} = \del^{\mu}_{\nu} ,
\hspace{2cm}
\left\{ \ps^a , \ps^b \right\} = -i \eta^{ab} ,
\label{2.6}
\ee

\nit
where $P_{\nu}$ is the momentum conjugate to $x^{\mu}$

\be
P_{\mu} = \dot{x}^{\nu} g_{\mu \nu}
- \frac{i}{2} \ps^a \ps^b \left( \og_{\mu a b} + A_{\mu a b} \right) + A_{\mu}
{}.
\label{2.7}
\ee

\nit
If, for notational convenience, we also define a (modified) covariant momentum
$\Pi_{\mu}$ ($\tilde{\Pi}_{\mu}$) by

\be
\ba{lll}
\Pi_{\mu} & = & \dsp{P_{\mu}
+ \frac{i}{2} \ps^a \ps^b \left( \og_{\mu a b} + A_{\mu a b} \right)
- A_{\mu}} , \\
& & \\
\tilde{\Pi}_{\mu} & = & \dsp{P_{\mu}
+ \frac{i}{2} \ps^a \ps^b \left( \og_{\mu a b} + \frac{1}{3}
A_{\mu a b} \right) - A_{\mu}} ,
\ea
\label{2.8}
\ee

\nit
the supercharge $Q$ and the Hamiltonian $H$ are given by

\be
\ba{lll}
Q & = & \ps^a e_{a}^{\;\; \mu} \tilde{\Pi}_{\mu} , \\
& & \\
H & = & \dsp{\frac{1}{2} \Pi_{\mu} \Pi_{\nu} g^{\mu \nu}
+ \frac{i}{2} \ps^a \ps^b F_{a b}
+ \frac{1}{4!} \ps^a \ps^b \ps^c \ps^d F_{a b c d}} .
\ea
\label{2.9}
\ee

\nit
Using the brackets given by eq.(\ref{2.6}) one can check that $Q$ is indeed
conserved

\be
\left\{ H , Q \right\} = 0 ,
\label{2.10}
\ee

\nit
and that

\be
\left\{ Q , Q \right\} = -2iH .
\label{2.11}
\ee

\nit
Eq.(\ref{2.11}) implies eq.(\ref{2.10}) by the Jacobi-identities.

The Poisson-Dirac brackets for general phase-space functions
$F(x,\Pi, \psi)$ can be cast in a manifest covariant form:

\be
\ba{lll}
\left\{ F, G \right\} & = & \dsp{ \cD_{\mu}^{T} F \dd{G}{\Pi_{\mu}}\, -\,
  \dd{F}{\Pi_{\mu}} \cD_{\mu}^{T} G } \\
  & & \\
  & + & \dsp{ \lh - \frac{i}{2} \psi^a \psi^b R_{ab\mu\nu}^{T}
      - 2 A_{\mu\nu}^{\:\:\:\:\:\lb} \Pi_{\lb} + F_{\mu\nu} \rh\,
      \dd{F}{\Pi_{\mu}} \dd{G}{\Pi_{\nu}}\, +\,
      i (-1)^{a_F} \dd{F}{\psi^a} \dd{G}{\psi^b}. }
\ea
\label{2.12}
\ee

\nit
with the covariant derivatives defined as

\be
\cD_{\mu}^{T} F\, =\, \pl_{\mu} F\, +\, \lh \og_{\mu ab} + A_{\mu ab} \rh
 \psi^b \dd{F}{\psi_a}\, +\, \lh \Gam_{\mu\nu}^{\:\:\:\:\:\lb} +
   A_{\mu \nu}^{\:\:\:\:\:\lb} \rh \Pi_{\lb} \dd{F}{\Pi_{\nu}}.
\label{2.13}
\ee

\nit
and with $a_{F}$ the Grassmann parity of $F$: $a_{F} = (0,1)$ for $F =
($even,odd$)$. As before the superscript $T$ denotes the inclusion of torsion.
Note that the brackets satisfy the Ricci identity in the presence of torsion
and electro-magnetism:

\be
\left\{ \Pi_{\mu}, \Pi_{\nu} \right\}\, =\, - \frac{i}{2} \psi^a \psi^b
    R_{ab\mu\nu}^{T} - 2 A_{\mu\nu}^{\:\:\:\:\:\lb} \Pi_{\lb} + F_{\mu\nu}.
\label{2.14}
\ee

\nit
The covariant expression (\ref{2.12}) for the brackets simplifies many
calculations considerably.

\section{New supersymmetries}{\label{S.5}}

The spinning particle model was constructed to be supersymmetric. Therefore,
independent of the form of the metric there is always a conserved supercharge
given by eq.(\ref{2.9}). Of course it is possible that the model has more
symmetry, but this will in general depend on the metric. In \ct{RvH3} it was
found that the model admits an extra, generalised type of supersymmetry when
the metric admits a tensor $f_{\mu \nu}$ satisfying

\be
f_{\mu \nu} = - f_{\nu \mu} ,
\label{3.1}
\ee

\be
f_{\mu \kg ; \nu} + f_{\nu \kg ; \mu} = 0 .
\label{3.2}
\ee

\nit
Such a tensor is called a Killing-Yano tensor. We will explain how this result
was obtained and how it generalises to the case where electro-magnetism and
torsion are present. Consider a quantity $Q_f$ of the form

\be
Q_f = \ps^a f_{a}^{\; \mu} \tilde{\Pi}_{\mu}
+ \frac{i}{3!} \ps^a \ps^b \ps^c c_{a b c} .
\label{3.3}
\ee

\nit
This quantity is invariant under supersymmetry if

\be
\left\{ Q , Q_f \right\} = 0 .
\label{3.4}
\ee

\nit
The Jacobi identities then guarantee that it is also conserved

\be
\left\{ H , Q_f \right\} =
0,
\label{3.5}
\ee

\nit
and hence it generates a symmetry of the action. Condition (\ref{3.4})
imposes constraints on $f_{a}^{\; \mu}$ and $c_{a b c}$. They read

\be
e_{a}^{\;\; (\mu} f_{b}^{\;\; \nu)} \eta^{a b} = 0 ,
\label{3.6}
\ee

\be
D_{\mu}^{T} f_{\nu}^{\;\; a} + D_{\nu}^{T} f_{\mu}^{\;\; a} = 0 ,
\label{3.7}
\ee

\be
e_{[a}^{\;\;\; \mu} f_{b]}^{\;\; \nu} F_{\mu \nu} = 0 ,
\label{3.8}
\ee

\be
c_{a b c} = 2 e_{[a}^{\;\;\; \mu} e_{b|}^{\;\; \nu} D_{\nu} f_{|c] \mu} .
\label{3.9}
\ee

\nit
Here $D_{\mu}$ denotes a derivative without torsion, while $D_{\mu}^{T}$ is a
derivative with torsion. Eq.(\ref{3.6}) says that $f^{\mu \nu}
:= e^{a \mu} f_{a}^{\;\; \nu}$ is anti-symmetric. Eq.(\ref{3.7})
generalises the Killing-Yano equation (\ref{3.2}) to the case with torsion.
When there is an electro-magnetic field, there is an extra condition,
eq.(\ref{3.8}), on $f_{a}^{\;\; \mu}$ to define a conserved quantity $Q_f$.
In \ct{RvH3}, where the Killing-Yano tensor of the Kerr-Newman metric was
considered, this last relation was not taken into account. However, it has
been checked now that it is satisfied in that case. Once a tensor has been
found which satisfies these three conditions, the tensor $c_{a b c}$ is
defined by eq.(\ref{3.9}) and a new Grassmann-odd constant of motion is
obtained. This charge is not necessarily a second supercharge since in
general

\be
\left\{ Q_f , Q_f \right\} = -2iZ \neq -2iH .
\label{3.10}
\ee

\nit
Hence we refer to it as a generalised supercharge. It is straightforward
to check, using the Jacobi identities, that the only non-vanishing brackets
between $Q,Q_f,H$ and $Z$ are (\ref{2.11}) and (\ref{3.10}).

\section{Dual spinning spaces}{\label{S.6}}

In sect.\ref{S.2} it was argued, that the most direct route to the duality
between metrics and Killing tensors comes from the symmetric Poisson-bracket
relation

\[ \left\{ H, K \right\}\, =\, 0. \]

\nit
The Poisson-Dirac bracket (\ref{3.4}) between ordinary and generalized
supercharges:

\[ \left\{ Q, Q_f \right\}\, =\, 0 \]

\nit
suggests a similar duality between the vielbein $e_{a}^{\:\: \mu}$ and
the Killing-Yano tensor $f_{a}^{\:\: \mu}$, interchanging ordinary and
generalized supersymmetry. In this section we show that this is indeed
the case, provided that we allow explicitly for the presence of torsion.

Consider two spinning particle models, labeled $I$ and $II$ respectively,
both describing a spinning particle interacting with gravitational,
electro-magnetic and torsion fields. Let theory $I$ also admit a Killing-Yano
tensor $f_{a}^{\;\; \mu}$ satisfying eqs.(\ref{3.6}) - (\ref{3.8}),
and let

\be
f_{a}^{\;\; \mu} f_{b}^{\;\; \nu} \eta^{a b} = K^{\mu \nu} .
\label{4.1}
\ee

\nit
The symmetric tensor $K$ on the right-hand side is a second-rank
Killing tensor \ct{RvH3}. Hence if $K_{\mu\nu} \neq g_{\mu\nu}$,
the metric of this space has a dual in the sense of sect.\ref{S.2}.
Note that in the present class of models there also exist examples
in which $K_{\mu\nu} = g_{\mu\nu}$, even though $Q_f \neq Q$. This
case corresponds to $N$-extended supersymmetry $(N \geq 2)$. We might
say that $N$-extended supersymmetric theories are self-dual in our
geometric sense.

The spinning-particle theory is completely determined by its supercharge $Q$,
as given by eq.(\ref{2.9}). The term in $Q$ proportional to $P_{\mu}$ defines
the vielbein and thus the metric. This determines the spin connection, which
implies that the torsion can be calculated from the terms proportional to
$\ps^a \ps^b \ps^c$. Finally the electro-magnetic field is found from the term
proportional to $\ps^a$.

Let us first introduce some notation. Theory $I$ is defined by the vielbein
$e_{a}^{\;\; \mu}$, the anti-symmetric torsion field $A_{abc}$ and the
electro-magnetic field $A_{\mu}$. The momentum $P_{\mu}$ is given by
eq.(\ref{2.7}). For general functions $F$ and $G$ of the phase-space
variables  $(x,P,\ps)$ the non-covariant form of the Poisson-Dirac bracket
reads

\be
\left\{ F , G \right\}_{I} =
\dd{F}{x^{\mu}} \dd{G}{P_{\mu}} - \dd{F}{P_{\mu}} \dd{G}{x^{\mu}}
+ i (-1)^{a_{F}} \dd{F}{\ps^a} \dd{G}{\ps_a} .
\label{4.2}
\ee

\nit
The theory has a supercharge $Q$ and a generalised supercharge $Q_f$ defined
by

\be
Q = \ps^a e_{a}^{\;\; \mu} P_{\mu}
+ \frac{i}{2} \ps^a \ps^b \ps^c
\left( e_{a}^{\;\; \mu} \og_{\mu bc} + \frac{1}{3} A_{abc} \right)
- \ps^a e_{a}^{\;\; \mu} A_{\mu} ,
\label{4.3}
\ee

\be
Q_f = \ps^a f_{a}^{\;\; \mu} P_{\mu}
+ \frac{i}{2} \ps^a \ps^b \ps^c
\left( f_{a}^{\;\; \mu} \og_{\mu bc}
+ \frac{1}{3} f_{a}^{\;\; \mu} e_{\mu}^{\;\; d} A_{dbc}
+ \frac{1}{3} c_{abc} \right)
- \ps^a f_{a}^{\;\; \mu} A_{\mu} ,
\label{4.4}
\ee

\nit
with

\be
\left\{ Q , Q_f \right\}_{I} = 0 .
\label{4.5}
\ee

\nit
The tensor $c_{abc}$ is given by eq.(\ref{3.9}).

Another spinning particle model, theory $II$, is defined by a vielbein
$\tilde{e}_{a}^{\;\; \mu}$, an anti-symmetric torsion field $\tilde{A}_{abc}$
and an electro-magnetic field $\tilde{A}_{\mu}$. The momentum $\bar{P}_{\mu}$
is given by

\be
\bar{P}_{\mu} = \dot{x}^{\nu} \tilde{g}_{\mu \nu}
- \frac{i}{2} \ps^a \ps^b \left( \tilde{\og}_{\mu ab} + \tilde{A}_{\mu ab}
\right) + \tilde{A}_{\mu} .
\label{4.6}
\ee

\nit
Here $\tilde{g}_{\mu \nu}$ and $\tilde{\og}_{\mu ab}$ are the metric and
spin connection calculated from $\tilde{e}_{a}^{\;\; \mu}$ in a way analogous
to eqs.(\ref{2.2}) and (\ref{2.3}). Since $\bar{P}_{\mu} \neq P_{\mu}$, theory
$II$ has different Dirac brackets

\be
\left\{ F , G \right\}_{II} =
\dd{F}{x^{\mu}} \dd{G}{\bar{P}_{\mu}} - \dd{F}{\bar{P}_{\mu}} \dd{G}{x^{\mu}}
+ i (-1)^{a_{F}} \dd{F}{\ps^a} \dd{G}{\ps_a} .
\label{4.7}
\ee

\nit
However, if we define an operation $F \rightarrow \bar{F}$ by

\be
F(x,P,\ps) \rightarrow \overline{F(x,P,\ps)} = F(x,\bar{P},\ps) ,
\label{4.8}
\ee

\nit
it is straightforward to deduce

\be
\left\{ \bar{F} , \bar{G} \right\}_{II}
= \overline{\left\{ F , G \right\}_{I}} .
\label{4.9}
\ee

\nit
Theory $II$ has a supercharge as well. It is given by

\be
\tilde{Q}  = \ps^a \tilde{e}_{a}^{\;\; \mu} \bar{P}_{\mu}
+ \frac{i}{2} \ps^a \ps^b \ps^c
\left( \tilde{e}_{a}^{\;\; \mu} \tilde{\og}_{\mu b c}
+ \frac{1}{3} \tilde{A}_{a b c} \right)
- \ps^a \tilde{e}_{a}^{\;\; \mu} \tilde{A}_{\mu} ,
\label{4.10}
\ee

\nit
We call theories $I$ and $II$ dual to each other if

\be
\tilde{e}_{a}^{\;\; \mu} = f_{a}^{\;\; \mu}.
\label{4.11}
\ee

\nit
We then want to know whether there is a Killing-Yano tensor

\be
\tilde{f}_{a}^{\;\; \mu} = e_{a}^{\;\; \mu} ,
\label{4.12}
\ee

\nit
for theory $II$, defining a generalised supercharge for that
theory\footnote{Notice that the transformation $\tilde{e} = f$ and
$\tilde{f} = e$ is quite subtle:

\be \nonumber
\ba{lllllll}
\tilde{e}_{a}^{\;\; \mu} \!\!\! & = & \!\!\! f_{a}^{\;\; \mu} & \Rightarrow
& \tilde{e}_{\mu}^{\;\; a} \!\!\! & =  & \!\!\!
\left( f^{-1} \right)_{\mu}^{\;\; a} , \\
\tilde{f}_{a}^{\;\; \mu} \!\!\! & = & \!\!\! e_{a}^{\;\; \mu} & \Rightarrow
& \tilde{f}_{\mu}^{\;\; a}
\!\!\! & = & \!\!\! \tilde{e}_{\mu}^{\;\; b} \tilde{f}_{b}^{\;\; \nu}
\tilde{e}_{\nu}^{\;\; a} \; = \; \left( f^{-1} \right)_{\mu}^{\;\; b}
e_{b}^{\;\; \nu} \left( f^{-1} \right)_{\nu}^{\;\; a} .
\ea
\ee}:

\be
\tilde{Q}_f = \ps^a \tilde{f}_{a}^{\;\; \mu} \bar{P}_{\mu}
+ \frac{i}{2} \ps^a \ps^b \ps^c
\left( \tilde{f}_{a}^{\;\; \mu} \tilde{\og}_{\mu b c}
+ \frac{1}{3} \tilde{f}_{a}^{\;\; \mu} \tilde{e}_{\mu}^{\;\; d} \tilde{A}_{dbc}
+ \frac{1}{3} \tilde{c}_{abc} \right)
- \ps^a \tilde{f}_{a}^{\;\; \mu} \tilde{A}_{\mu}
\label{4.13}
\ee

\nit
(with $\tilde{c}_{abc}$ defined analogous to eq.(\ref{3.9})). It should
satisfy the relation

\be
\left\{ \tilde{Q} , \tilde{Q}_f \right\}_{II} = 0 .
\label{4.14}
\ee

\nit
Now eqs.(\ref{4.13}), (\ref{4.14}) are obtained if one takes

\be
\tilde{Q} = \overline{Q_f} ,
\label{4.15}
\ee

\be
\tilde{Q}_f = \overline{Q} .
\label{4.16}
\ee

\nit
Indeed, eqs.(\ref{4.9}) and (\ref{4.5}) then immediately lead to the result

\be
\left\{ \tilde{Q} , \tilde{Q}_f \right\}_{II} =
\left\{ \overline{Q_f} , \overline{Q} \right\}_{II} =
\overline{\left\{ Q_f , Q \right\}_{I}} = 0 .
\label{4.17}
\ee

\nit
Comparing eqs.(\ref{4.4}) and (\ref{4.10}) one finds the conditions

\be
\tilde{e}_{a}^{\;\; \mu} = f_{a}^{\;\; \mu} ,
\label{4.18}
\ee

\be
3 \tilde{e}_{[a|}^{\;\; \mu} \tilde{\og}_{\mu|bc]} + \tilde{A}_{abc} =
3 f_{[a|}^{\;\; \mu} \og_{\mu|bc]} +
f_{[a|}^{\;\; \mu} e_{\mu}^{\;\; d} A_{d|bc]} + c_{abc} ,
\label{4.19}
\ee

\be
\tilde{A}_{\mu} = A_{\mu} ,
\label{4.20}
\ee

\nit
while comparison of eqs.(\ref{4.3}) and (\ref{4.13}) gives

\be
\tilde{f}_{a}^{\;\; \mu} = e_{a}^{\;\; \mu} ,
\label{4.21}
\ee

\be
3 \tilde{f}_{[a|}^{\;\; \mu} \tilde{\og}_{\mu|bc]} +
\tilde{f}_{[a|}^{\;\; \mu} \tilde{e}_{\mu}^{\;\; d} \tilde{A}_{d|bc]}
+ \tilde{c}_{abc} =
3 e_{[a|}^{\;\; \mu} \og_{\mu|bc]} + A_{abc} ,
\label{4.22}
\ee

\be
\tilde{A}_{\mu} = A_{\mu} ,
\label{4.23}
\ee

\nit
Eqs.(\ref{4.18}) and (\ref{4.21}) are satisfied by definition and
eqs.(\ref{4.20}) and (\ref{4.23}) are consistent. Eq.(\ref{4.19}) defines
$\tilde{A}_{abc}$ in terms of the background fields of theory $I$ and
$\tilde{\og}_{\mu a b}$. This last quantity is fixed once
$\tilde{e}_{a}^{\;\; \mu}$ is fixed by eq.(\ref{4.18}). Hence we find

\be
\tilde{A}_{abc} = -3 f_{[a|}^{\;\; \mu}
\left( \tilde{\og}_{\mu |bc]} - \og_{\mu |bc]} \right)
+ f_{[a|}^{\;\; \mu} e_{\mu}^{\;\; d} A_{d|bc]} + c_{abc}.
\label{4.24}
\ee

\nit
Eq.(\ref{4.24}) then defines $\tilde{c}_{abc}$ as

\be
\tilde{c}_{abc}
= -3 e_{[a|}^{\;\; \mu} \left( \tilde{\og}_{\mu |bc]} - \og_{\mu |bc]} \right)
- e_{[a|}^{\;\; \mu} \left( f^{-1} \right)_{\mu}^{\;\; d} \tilde{A}_{d|bc]}
+ A_{abc} .
\label{4.25}
\ee

\nit
Hence there is always an extra supercharge $\tilde{Q}$, defined as given
above, for theory $II$ such that eq.(\ref{4.14}) is satisfied.

One can also show this more explicitly. Working out eq.(\ref{4.14}) one finds
equations similar to (\ref{3.6}) - (\ref{3.9}):

\be
\tilde{e}_{a}^{\;\; (\mu} \tilde{f}_{b}^{\;\; \nu)} \eta^{a b} = 0 ,
\label{4.26}
\ee

\be
\tilde{D}_{\mu}^{T} \tilde{f}_{\nu}^{\;\; a} +
\tilde{D}_{\nu}^{T} \tilde{f}_{\mu}^{\;\; a} = 0 ,
\label{4.27}
\ee

\be
\tilde{e}_{[a}^{\;\;\; \mu} \tilde{f}_{b]}^{\;\; \nu} \tilde{F}_{\mu \nu} = 0 ,
\label{4.28}
\ee

\be
\tilde{c}_{a b c} = 2 \tilde{e}_{[a}^{\;\;\; \mu} \tilde{e}_{b|}^{\;\; \nu}
\tilde{D}_{\nu} \tilde{f}_{|c] \mu} .
\label{4.29}
\ee

\nit
Eqs.(\ref{4.26}) and (\ref{4.28}) are obviously equal to their dual versions in
theory $I$ and therefore automatically satisfied. Eqs.(\ref{4.27}) and
(\ref{4.29}) are less straightforward to check but, using eqs.(\ref{3.6}) -
(\ref{3.9}), one can indeed show that they are satisfied. Eq.(\ref{4.27}) shows
that theory $II$ also has a Killing-Yano tensor.

Hence we have shown that a theory with a Killing-Yano tensor always has a
dual theory in which the vielbein and the Killing-Yano tensor have reversed
r\^{o}les. Eqs.(\ref{4.20}) and (\ref{4.24}) then prescribe what the
electro-magnetic field and the torsion of the dual theory are.
Notice that the introduction of torsion is crucial here. A theory without
torsion will have a dual theory with torsion in general.

\section{Kerr-Newman and Taub-NUT}{\label{S.7}}

Again we illustrate the general theorem derived above with the
examples of the geometries of Kerr-Newman \ct{RvH3} and Taub-NUT \ct{Vis,JW}.
\vspace{2ex}

\nit
{\em Kerr-Newman}. First consider standard Kerr-Newman space-time, specified
by the metric (\ref{1.12}) and the Maxwell 2-form (\ref{1.15}). In this case
the torsion $A_{abc}$ vanishes. It was found by Penrose and Floyd \ct{PF}
that the Kerr-Newman metric admits a Killing-Yano tensor and hence there is
an extra supersymmetry. This was described in \ct{RvH3} and we will give the
results below.

The Kerr-Newman metric can be described by the following vierbein
components:

\be
\ba{lll}
e_{0}^{\;\; \mu} \partial_{\mu} & = &
\dsp{- \frac{1}{\rho \sqrt{\Delta}} \left[ \left( r^2 + a^2 \right)
\partial_t + a \partial_{\phi} \right]} , \\
& & \\
e_{1}^{\;\; \mu} \partial_{\mu} & = &
\dsp{\frac{\sqrt{\Delta}}{\rho} \partial_r} , \\
& & \\
e_{2}^{\;\; \mu} \partial_{\mu} & = &
\dsp{\frac{1}{\rho} \partial_{\theta}} , \\
& & \\
e_{3}^{\;\; \mu} \partial_{\mu} & = &
\dsp{\frac{1}{\rho \sin \theta} \left[ a \sin^2 \theta \partial_t +
\partial_{\phi} \right]} .
\ea
\label{5.5}
\ee

\nit
For this geometry a Killing-Yano tensor exists, defined by

\be
\ba{lll}
f_{0}^{\;\; \mu} \partial_{\mu} & = &
\dsp{\frac{a \sqrt{\Delta} \cos \theta}{\rho} \partial_r} , \\
& & \\
f_{1}^{\;\; \mu} \partial_{\mu} & = &
\dsp{- \frac{a \cos \theta}{\rho \sqrt{\Delta}}
\left[ \left( r^2 + a^2 \right) \partial_t +
a \partial_{\phi} \right]} , \\
& & \\
f_{2}^{\;\; \mu} \partial_{\mu} & = &
\dsp{\frac{r}{\rho \sin \theta} \left[ a \sin^2 \theta \partial_t +
\partial_{\phi} \right]} , \\
& & \\
f_{3}^{\;\; \mu} \partial_{\mu} & = &
\dsp{- \frac{r}{\rho} \partial_{\theta}} .
\ea
\label{5.6}
\ee

\nit
This Killing-Yano tensor defines an extra supercharge as given by
eq.(\ref{3.4}) with

\be
\ba{lllllll}
\dsp{c_{0 1 2}} & = & \dsp{- \frac{2 a \sin \th}{\rg}} & \hspace{2cm}
 & \dsp{c_{0 1 3}} & = & 0 \\
 & & & & & & \\
\dsp{c_{0 2 3}} & = & 0 &
 & \dsp{c_{1 2 3}} & = & \dsp{\frac{2 \sqrt{\Del}}{\rg}}
\ea
\label{5.7}
\ee

\nit
If we take $f_{a}^{\mu}$ to be the vierbein of the dual theory, then we find
the dual metric given in eq.(\ref{1.17}), in combination with a non-vanishing
torsion

\be
\ba{lllllll}
\dsp{\tilde{A}_{0 1 2}} & = & \dsp{\frac{a \sin \th}{\rg}} & \hspace{2cm}
 & \dsp{\tilde{A}_{0 1 3}} & = & 0 \\
 & & & & & & \\
\dsp{\tilde{A}_{0 2 3}} & = & 0 &
 & \dsp{\tilde{A}_{1 2 3}} & = & \dsp{\frac{- \sqrt{\Del}}{\rg}}
\ea
\label{5.9}
\ee

\nit
In accordance with eq.(\ref{4.23}), the electro-magnetic field is the same
as in the original theory. Notice that in the dual theory $dt - a \sin^2 \th
d\fg$ has become space-like, while $r$ has become time-like.

In this dual theory the original vielbein $e_{a}^{\mu}$ is a Killing-Yano
tensor, defining an extra supersymmetry with non-vanishing anti-symmetric
third-rank tensors

\be
\ba{lllllll}
\dsp{\tilde{c}_{0 1 2}} & = & 0, &  &
\dsp{\tilde{c}_{0 1 3}} & = &
\dsp{- \frac{2(3r^2 - 2a^2 \cos^2 \theta) \sin \theta}{3 \rho ra \cos^2
\theta}, } \\
 & & & & & & \\
\dsp{\tilde{c}_{0 2 3}} & = &
\dsp{- \frac{2 \sqrt{\Del}(2r^2 - 3a^2 \cos^2 \theta)}{3\rho r^2 a \cos
\theta}, }
& \hspace{1cm}
 & \dsp{\tilde{c}_{1 2 3}} & = & 0.
\ea
\label{5.10}
\ee

\nit
{\em Taub-NUT}. Again we start from the standard form of the metric,
eq.(\ref{1.18}), without torsion. This metric can be described by the
following vielbein

\be
\ba{lll}
e_{0}^{\;\; \mu} \partial_{\mu} & = &
\dsp{\frac{1}{\sqrt{1 + \frac{2m}{r}}} \partial_r} , \\
& & \\
e_{1}^{\;\; \mu} \partial_{\mu} & = &
\dsp{\frac{1}{r \sqrt{1 + \frac{2m}{r}}} \partial_{\theta}} , \\
& & \\
e_{2}^{\;\; \mu} \partial_{\mu} & = &
\dsp{\frac{1}{r \sin \theta \sqrt{1 + \frac{2m}{r}}}
\left[ \partial_{\phi} - \cos \theta \; \partial_{\psi} \right]} , \\
& & \\
e_{3}^{\;\; \mu} \partial_{\mu} & = &
\dsp{\frac{\sqrt{1 + \frac{2m}{r}}}{2m} \partial_{\psi}} .
\ea
\label{5.111}
\ee

\nit
The metric allows four Killing-Yano tensors \ct{GRub}, but three of these are
trivial in the sense that they correspond to $N$-extended supersymmetry
\ct{JW}. They are defined by

\be
e_{a}^{(i)\mu} \; = \; M_{a}^{(i) b} e_{b}^{(0)\mu},
\hspace{1cm} (i=1,2,3),
\label{5.12}
\ee

\nit
where the matrices $M^{(i)}$ are of the form

\be
\ba{ccc}
M^{(1)} & = &
\left(
\ba{cccc}
0 & \sin \fg & \cos \th \cos \fg & - \sin \th \cos \fg \\
- \sin \fg & 0 & - \sin \th \cos \fg & - \cos \th \cos \fg \\
- \cos \th \cos \fg & \sin \th \cos \fg & 0 & \sin \fg \\
\sin \th \cos \fg & \cos \th \cos \fg & - \sin \fg & 0
\ea
\right), \\
& & \\
M^{(2)} & = &
\left(
\ba{cccc}
0 & - \cos \fg & \cos \th \sin \fg & - \sin \th \sin \fg \\
\cos \fg & 0 & - \sin \th \sin \fg & - \cos \th \sin \fg \\
- \cos \th \sin \fg & \sin \th \sin \fg & 0 & - \cos \fg \\
\sin \th \sin \fg & \cos \th \sin \fg &  \cos \fg & 0
\ea
\right), \\
& & \\
M^{(3)} & = &
\left(
\ba{cccc}
0 & 0 & - \sin \th & - \cos \th \\
0 & 0 & - \cos \th & \sin \th \\
\sin \th & \cos \th & 0 & 0 \\
\cos \th & - \sin \th & 0 & 0
\ea
\right), \\
\ea
\label{5.13}
\ee

\nit
satisfying the quaternion algebra

\be
M^{(i)} M^{(j)} \; = \; \eps^{ijk} M^{(k)} - \delta^{ij} \; 1.
\label{5.13.1}
\ee

\nit
The tensors $e_{a}^{(i)\mu}$ are covarianly constant and thus obey the
Killing-Yano equation in a trivial way. We indicate these Killing-Yano tensors
by an $e$ in stead of an $f$ because they define the same metric and thus
the same space. We label the four different representations of this space
defined by the four different vielbeine with $I=(0,i)=(0,1,2,3)$. One can
check explicitly that

\be
D^{(I)}_{\mu} e_{a}^{(J)\nu} \, = \, 0, \hspace{2cm} for \;\; I,J = 0,1,2,3
\label{5.14}
\ee

\nit
where $D^{(I)}_{\mu}$ is the covariant derivative in space $I$. Hence the
$e_{a}^{(J)\mu}$ play the r\^{o}le of (trivial) Killing-Yano tensors in all
representations $I$ of Taub-NUT space. It is found from eq.(\ref{3.9}) that
the corresponding tensors $c^{(I,J)}_{a b c}$ vanish for all $J$. Here we
introduced a notation where the first upper index labels the space, while the
second one labels the Killing-Yano tensor from which $c^{(I,J)}_{a b c}$
is computed ($J$ for $e_{a}^{(J)\mu}$).

Taub-NUT space admits a fourth Killing-Yano tensor $f_{a}^{\;\; \mu}$, which
is non-trivial in this respect. In space $I$ it is found from

\be
e^{(I) \mu}_{a}\, f_{a}^{(I) \nu} \, = \, f^{\mu \nu},
\label{5.15}
\ee

\nit
(no sum over $I$) where the anti-symmetric $f$-symbol $f^{\mu \nu}$ has
the explicit form

\be
f^{\mu \nu} \pl_{\mu} \wedge \pl_{\nu} \, = \,
- \frac{1}{m} \pl_{r} \wedge \pl_{\psi}
+ \frac{2(r+m)}{rm(r+2m) \sin \th} \pl_{\th} \wedge
\left( \pl_{\fg} - \cos \th \pl_{\psi} \right).
\label{5.16}
\ee

\nit
The symmetric Killing tensor (\ref{1.19}) is the square of these
Killing-Yano tensors and defines for all $I$ the {\em same} dual metric
(\ref{1.20}).

The corresponding anti-symmetric third-rank tensors are found from

\be
c_{a b c}^{(I,\tilde{I})} \, = \,
e^{(I) \mu}_{a} e^{(I) \nu}_{b} e^{(I) \kg}_{c} c_{\mu \nu \kg}.
\label{5.17}
\ee

\nit
Again the second upper index refers to the corresponding Killing-Yano
tensor: $(I,\tilde{J})$ labels the anti-symmetric three-tensor in space $I$
computed from $f_{a}^{(J) \mu}$. The tensor $c_{\mu \nu \kg}$,
which is the same for all $I$, has only one non-vanishing component

\be
c_{r \th \fg} \, = \, \frac{2r(r+2m) \sin \th}{m}.
\label{5.18}
\ee

\nit
We now investigate the duality structure of this theory. The contractions

\be
f^{(I,J) \mu \nu} \, = \, e_{a}^{(I) \mu} e_{a}^{(J) \nu}
\label{5.19}
\ee

\nit
are anti-symmetric for all $I \neq J$. Together with the fact that the
$e_{a}^{(I)\mu}$ are Killing-Yano tensors for space $I \neq J$ this implies
that spaces $I$ and $J$ are dual to each other. The torsion in space $J$ can
then be calculated from the torsion in space $I$ using eq.(\ref{4.24})

\be
A^{(J)}_{abc} \, = \, -3 e_{[a|}^{(J) \mu}
\left( \og^{(J)}_{\mu |bc]} - \og^{(I)}_{\mu |bc]} \right)
+ e_{[a|}^{(J) \mu}
\left( e^{(I)-1} \right)_{\mu}^{\;\; d} A^{(I)}_{d|bc]} + c^{(I,J)}_{abc}.
\label{5.20}
\ee

\nit
We found already that the tensors $c^{(I,J)}_{abc}$ vanish.
It is also straightforward to show, using eq.(\ref{5.14}) for different values
of $I$, that the spin connections $\og^{(I)}_{\mu a b}$ are all identical.
This implies that $A^{(J)}_{abc}$ is proportional to $A^{(I)}_{abc}$ for all
$I$ and $J$. Since the torsion vanishes in the space labeled (0), it vanishes
in all spaces $I$, which is consistent.

Now we turn to the duality between space $I$ and the space defined by the
vielbein $f_{a}^{(I)\mu}$. We will label this space by $\tilde{I}$. The
contraction $e_{a}^{(I)\mu} f_{a}^{(I)\nu}$ is anti-symmetric by definition
(\ref{5.16}) and furthermore $f_{a}^{(I)\mu}$ satisfies the Killing-Yano
equation for space $I$. Hence spaces $I$ and $\tilde{I}$ are dual to each
other and $e_{a}^{(I)\mu}$ must satisfy the Killing-Yano equation in space
$\tilde{I}$. The torsion in space $\tilde{I}$ is found to be defined by

\be
A_{a b c}^{(\tilde{I})} \, = \,
f^{(I) \mu}_{a} f^{(I) \nu}_{b} f^{(I) \kg}_{c} \tilde{A}_{\mu \nu \kg},
\label{5.21}
\ee

\nit
with $\tilde{A}_{\mu \nu \kg}$ having only one non-vanishing component

\be
\tilde{A}_{\th \fg \psi} \, = \, - \frac{2r^{2} m^{2} \sin \th}{(r+m)^{2}}.
\label{5.22}
\ee

\nit
Notice that $\tilde{A}_{\mu \nu \kg}$ is the same for all spaces $\tilde{I}$.
The conclusion is that there is only one space dual to Taub-NUT space. The
spaces $\tilde{I}$ have the same metric and the same torsion and thus
represent the same geometry.

Using the expression for the torsion we have checked explicitly that
$e_{a}^{(I)\mu}$ satisfies the Killing-Yano equation for space $\tilde{I}$.
The corresponding anti-symmetric three-tensors (space $\tilde{I}$, Killing-Yano
tensor $e^{(I) \mu}_{a}$) are defined by

\be
c_{a b c}^{(\tilde{I},I)} \, = \,
f^{(I) \mu}_{a} f^{(I) \nu}_{b} f^{(I) \kg}_{c} \tilde{c}_{\mu \nu \kg},
\label{5.23}
\ee

\nit
with

\be
\tilde{c}_{r \th \fg} \, = \,
\frac{2rm(r+2m)(2r^{2} + 4rm + 5m^{2}) \sin \th}{3(r+m)^{4}},
\label{5.24}
\ee

\nit
being the only non-vanishing component of $\tilde{c}_{\mu \nu \kg}$, which is
the same for all representations $\tilde{I}$ of dual Taub-NUT space.

The duality between Taub-NUT space and dual Taub-NUT space is not complete.
The described procedure gives only one Killing-Yano tensor for dual Taub-NUT
space. The $f_{a}^{(I)\mu}$ are not solutions of the Killing-Yano equation
for space $\tilde{J}$ if $J \neq I$. And although it is easily checked that
$f_{a}^{(J)\mu}$ satisfies the Killing-Yano equation for space $I$, also for
$I \neq J$ (the covariant derivatives are the same in all spaces I), there is
no duality between the spaces $I$ and $\tilde{J}$ because $e_{a}^{(I)\mu}
f_{a}^{(J)\nu}$ is not anti-symmetric for $I \neq J$. Hence also the
$e_{a}^{(I)\mu}$ will not automatically be Killing-Yano tensors for space
$\tilde{J}$, $J \neq I$. We have checked this explicitly and found that this
is indeed not the case.

Finally we remark that in space $J$ both $e_{a}^{(I)\mu}$ and $f_{a}^{(J)\nu}$
are Killing-Yano tensors which implies, as was shown in \ct{RvH3}, that in
that space the symmetric contraction

\be
e_{a}^{(I)\mu} f_{a}^{(J)\nu} + e_{a}^{(I)\nu} f_{a}^{(J)\mu}
\label{5.25}
\ee

\nit
defines a second rank Killing tensor. For $I \neq J$ these tensors are
non-trivial and are defined precisely by the conserved quantities $K^{(i)}$
given in (\ref{1.21}) for the scalar particle. We conclude that these second
rank Killing tensors are of a different type than the ones given by
(\ref{1.16}) and (\ref{1.19}). Although they do define a dual
bosonic space, they donot define a dual spinning space, since they are not
the square of a Killing-Yano tensor.

\vspace{5mm}

\nit
{\bf Acknowledgement}

\nit
One of us (RHR) would like to thank the Engineering and Physical Sciences
Research Council for post-doctoral support and Koos de Vos for very useful and
pleasant conversations.

\end{document}